# Iteratively-Reweighted Least-Squares Fitting of Support Vector Machines: A Majorization–Minimization Algorithm Approach

Hien D. Nguyen* and Geoffrey J. McLachlan

May 12, 2017

**Abstract**

Support vector machines (SVMs) are an important tool in modern data analysis. Traditionally, support vector machines have been fitted via quadratic programming, either using purpose-built or off-the-shelf algorithms. We present an alternative approach to SVM fitting via the majorization–minimization (MM) paradigm. Algorithms that are derived via MM algorithm constructions can be shown to monotonically decrease their objectives at each iteration, as well as be globally convergent to stationary points. We demonstrate the construction of iteratively-reweighted least-squares (IRLS) algorithms, via the MM paradigm, for SVM risk minimization problems involving the hinge, least-square, squared-hinge, and logistic losses, and 1-norm, 2-norm, and elastic net penalizations. Successful implementations of our algorithms are presented via some numerical examples.

*HDN is at the Department of Mathematics and Statistics, La Trobe University, Bundoora Victoria, Australia 3086 (email: h.nguyen5@latrobe.edu.au). GJM is at the School of Mathematics and Statistics, University of Queensland, St. Lucia Queensland, Australia 4072.

# 1 Introduction

Ever since their introduction by [3], support vector machines (SVMs) have become a mainstay in the toolkit of modern data analysts and machine learning practitioners. The popularity of SVMs is well-earned as they generally perform favorably when compared to other off-the-shelf classification techniques; see, for example, [25].

Let $(\boldsymbol{X}, Y) \in \mathbb{X} \times \{-1, 1\}$ be a random observation in some probability space (consisting of a feature vector $\boldsymbol{X}$ and a classification label $Y$), where $\mathbb{X} \subset \mathbb{R}^p$ for $p \in \mathbb{N}$. Suppose further that $\boldsymbol{T} : \mathbb{X} \to \mathbb{T}$ is a mapping of $\boldsymbol{X}$ from $\mathbb{X}$ into some transformation space $\mathbb{T} \subset \mathbb{R}^q$ for $q \in \mathbb{N}$. For a new observation, $\boldsymbol{T}(\boldsymbol{X})$, in the transformed feature space, let $\boldsymbol{Z}^\top = \left(\boldsymbol{T}^\top(\boldsymbol{X}), Y\right)$. Here, $\top$ indicates transposition.

When constructing a soft-margin binary SVM classifier, we wish to obtain some hyperplane $\alpha + \boldsymbol{\beta}^\top \boldsymbol{t} = 0$, such that there is a high occurrence probability of the event $Y\left[\alpha + \boldsymbol{T}^\top \boldsymbol{\beta}\right] > 0$, where $\alpha \in \mathbb{R}$, $\boldsymbol{\beta}^\top = (\beta_1, ..., \beta_q) \in \mathbb{R}^q$, and $\boldsymbol{z}^\top = (\boldsymbol{t}, y)$ is a fixed realization of $\boldsymbol{Z}$. We say that $\boldsymbol{\theta}^\top = \left(\alpha, \boldsymbol{\beta}^\top\right)$ is the parameter vector of the hyperplane.

Let $\mathbf{Z} = \{\boldsymbol{Z}_i\}_{i=1}^n$ be an IID (independent and identically distributed) sample consisting of $n \in \mathbb{N}$ observations. Under the empirical risk minimization framework of [26], the problem of obtaining an optimal hyperplane can be cast as the minimization problem

$$\hat{\boldsymbol{\theta}} = \arg\min_{\alpha, \boldsymbol{\beta}} \ R_n(\alpha, \boldsymbol{\beta}; \mathbf{Z}), \tag{1}$$

where $\hat{\boldsymbol{\theta}}^\top = \left(\hat{\alpha}, \hat{\boldsymbol{\beta}}^\top\right)$ is the optimal parameter vector. The risk $R_n$ can be expanded into its two components $n^{-1}\sum_{i=1}^{n} l\left(\alpha + \boldsymbol{\beta}^\top \boldsymbol{T}_i, Y_i\right) + P(\boldsymbol{\beta})$, where $n^{-1}\sum_{i=1}^{n} l\left(\alpha + \boldsymbol{\beta}^\top \boldsymbol{T}_i, Y_i\right)$ is the average loss and $P(\boldsymbol{\beta})$ is a penalization function.

Under different specifications of the SVM framework, the loss function $l$ and penalty function $P$ can take varying forms. In the original specification (i.e. in [3]), the loss and penalty are taken to be the hinge loss function $l(w, y) = [1 - wy]_+$ and the quadratic (2-norm) penalty $P(\boldsymbol{\beta}) = \lambda \boldsymbol{\beta}^\top \boldsymbol{\beta}$, respective, where $[w]_+ = \max\{0, w\}$ for $w \in \mathbb{R}$, and $\lambda > 0$ is some penalization constant. Alternatively, [24] suggested the use of the least-squares criterion $l(w, y) = (1 - wy)^2$, instead. Some other loss functions include the squared-hinge loss $l(w, y) = [1 - wy]_+^2$, and logistic loss $l(w, y) = \log[1 + \exp(-wy)]$ (cf. [30]).

Similarly, various alternatives to the quadratic penalty have been suggested. For example [31] considered the LASSO (1-norm) penalty $P(\boldsymbol{\beta}) = \mu \sum_{j=1}^{q} |\beta_j|$, where $\mu > 0$ is a penalization constant. Another alternative is the elastic net-type penalty $P(\boldsymbol{\beta}) = \lambda \boldsymbol{\beta}^\top \boldsymbol{\beta} + \mu \sum_{j=1}^{q} |\beta_j|$ of [27].

The conventional methodology for computing (1) is to state the problem as a quadratic program, and then to solve said program via some off-the-shelf or purpose-built algorithm. See, for example, the setup in Appendix 1 of [3] and the expanded exposition of [1, Ch. 5].

As an alternative to quadratic programming, [16] derived an iteratively-reweighted least-squares (IRLS) algorithm for computing (1) under the original choices of loss and penalty of [3], via careful manipulation of the KKT conditions (Karush-Kuhn-Tucker; cf [19, Ch. 16]). Using the obtained IRLS algorithm, [17] proposed a modification for the distributed fitting of SVMs across a network.

In [20], an attempt to generalize the result of [16] to arbitrary loss functions was proposed. However, due to the wide range of loss functions that was considered in the article (e.g. some losses were not convex), [20] were required to include a line search step in their proposed IRLS-type algorithm. Further, they were not able to provide any global convergence guarantees. In this article, we will restrict our attention to constructing IRLS algorithms for computing (1), where the risk consists of losses and penalties that are of the forms discussed above. Note that all of the losses and penalties are convex, and thus our scope is less ambitious, although more manageable than that of [20].

Using the MM (majorization–minimization) paradigm of [9] (see also [12]) and following suggestions from [4] and [28], an IRLS algorithm for computing (1) under the specification of [3] was proposed in [18]. In this article, we generalize the result of [18] by showing that one can compute (1) for risk functions that consist of any combination of losses and penalties from above via an IRLS algorithm by invoking the MM paradigm. Furthermore, we show that the constructed IRLS algorithms are all monotonic and globally convergent in the sense that the risk at every iteration is decreasing and that the iterates approach the global minimum of the risk. We refer the interested reader to [9] for a short tutorial on MM algorithms, and to [21] and [7] for an engineering perspective.

The algorithms that we present are illustrative of the MM paradigm and we do not suggest that the results that we present, in their current forms, are direct substitutes for the methods used in state-of-the-art solvers such as those of [22], [5], [6], and [11]; see also [23]. However, we believe that the new methods present an interesting perspective on the estimation of SVM classifiers that may be useful in combinations of losses and penalties where no current best-practice methods exist. Furthermore, we believe that further development and research into optimal implementations of MM based methods, as well as hybridizing MM methods with current approaches, may yield computational gains on the current state-of-the-art.

The article proceeds with an introduction to MM algorithms in Section 2. MM algorithms are constructed in Section 3. Numerical examples are presented in Section 4. Finally, conclusions are drawn in Section 5.

## 2 Introduction to MM Algorithms

Let $F(\boldsymbol{u})$ be some objective function of interest that one wishes to minimize, where $\boldsymbol{u} \in \mathbb{U} \subset \mathbb{R}^r$ for some $r \in \mathbb{N}$. Suppose, however, that $F$ is difficult to manipulate (e.g. non-differentiable or multimodal). Let $M(\boldsymbol{u}; \boldsymbol{v})$ be defined as a majorizer of $F$ at $\boldsymbol{v} \in \mathbb{U}$ if (i) $M(\boldsymbol{u}; \boldsymbol{u}) = F(\boldsymbol{u})$ for all $\boldsymbol{u}$, and (ii) $M(\boldsymbol{u}; \boldsymbol{v}) \geq F(\boldsymbol{u})$ for $\boldsymbol{u} \neq \boldsymbol{v}$.

Starting from some initial value $\boldsymbol{u}^{(0)}$, we say that $\{\boldsymbol{u}^{(k)}\}$ is a sequence of MM algorithm iterates for minimizing $F$ if it satisfies the condition

$$\boldsymbol{u}^{(k+1)} = \arg\min_{\boldsymbol{u} \in \mathbb{U}} M\left(\boldsymbol{u}; \boldsymbol{u}^{(k)}\right), \qquad (2)$$

for each $k \in \mathbb{N} \cup \{0\}$. The definition (2) guarantees the monotonicity of any MM algorithm.

**Proposition 1.** *Starting from some $\boldsymbol{u}^{(0)}$, if $\{\boldsymbol{u}^{(k)}\}$ is a sequence of MM algorithm iterates, then $\{F(\boldsymbol{u}^{(k)})\}$ is monotonically decreasing.*

*Proof.* For any $k$, the definition of a majorizer implies that

$$\begin{aligned} F\left(\boldsymbol{u}^{(k)}\right) &= M\left(\boldsymbol{u}^{(k)}; \boldsymbol{u}^{(k)}\right) \\ &\geq M\left(\boldsymbol{u}^{(k+1)}; \boldsymbol{u}^{(k)}\right) \\ &\geq F\left(\boldsymbol{u}^{(k+1)}\right), \end{aligned}$$

where the second line follows from (2). □

Define the directional derivative of $F$ at $\boldsymbol{u}$ in the direction $\boldsymbol{\delta}$ as

$$F'(\boldsymbol{u};\boldsymbol{\delta}) = \lim_{\lambda \downarrow 0} \frac{F(\boldsymbol{u}+\lambda\boldsymbol{\delta}) - F(\boldsymbol{u})}{\lambda},$$

and define a stationary point of $F$ to be any point $\boldsymbol{u}^*$ that satisfies the condition $F'(\boldsymbol{u}^*;\boldsymbol{\delta}) \geq 0$, for all $\boldsymbol{\delta}$ such that $\boldsymbol{u} + \boldsymbol{\delta}$ is a valid input of $F$.

In addition to the monotonicity property, it is known that MM algorithms are in globally convergent, in general, under some generous conditions. Let $\boldsymbol{u}^{(\infty)} = \lim_{k\to\infty} \boldsymbol{u}^{(k)}$ be the limit point of the MM algorithm, for some initial value $\boldsymbol{u}^{(0)}$. The following result is from [21].

**Proposition 2.** *Make the assumption that $M'(\boldsymbol{u},\boldsymbol{v};\boldsymbol{\delta})|_{\boldsymbol{u}=\boldsymbol{v}} = F'(\boldsymbol{v};\boldsymbol{\delta})$ for all $\boldsymbol{\delta}$ such that $\boldsymbol{u}+\boldsymbol{\delta}$ is a valid input of both $M$ and $F$, and that $M(\boldsymbol{u},\boldsymbol{v})$ is continuous in both coordinates. Starting from some $\boldsymbol{u}^{(0)}$, if $\boldsymbol{u}^{(\infty)}$ is the limit point of the MM algorithm iterates $\{\boldsymbol{u}^{(k)}\}$, then $\boldsymbol{u}^{(\infty)}$ is a stationary point of the objective function $F$. Further, if $F$ is convex, then $\boldsymbol{u}^{(\infty)}$ is a global minimum of $F$.*

There are numerous ways to construct majorizers; see the comprehensive treatment in [12, Ch. 4]. For the purpose of our exposition, we only require the following results.

**Lemma 3.** *Let $F(u) = f(u)$ be a twice differentiable function. If the second derivative satisfies $\Delta \geq f''(u)$ for all $u \in \mathbb{U}$, then $F$ is majorized at $v$ by*

$$M(u;v) = f(v) + f'(v)(u-v) + \frac{\Delta}{2}(u-v)^2.$$

**Lemma 4.** *For any $a \in [1,2]$, if $v \neq 0$, then $F(u) = |u|^a$ can be majorized at $v$ by*

$$M(u;v) = \frac{a}{2}|v|^{a-2}u^2 + \left(1 - \frac{a}{2}\right)|v|^a.$$

**Corollary 5.** *If $v \neq 0$, then the function $F(u) = [u]_+$ is majorized at $v$ by*

$$M(u; v) = \frac{1}{4|v|} (u + |v|)^2.$$

*Proof.* Start with the identity $\max\{a, b\} = |a - b|/2 + a/2 + b/2$ and substitute $a = u$ and $b = 0$ to get $F(u) = \max\{u, 0\} = [u]_+ = |u|/2 + u/2$. Apply Lemma 4 to $|u|$ (i.e. $a = 1$) to get $M(u; v) = (u^2/2|v| + |v|/2)/2 + u/2 = (u + |v|)^2/4|v|$, as required. $\square$

Lemma 3 is due to [2] and Lemma 4 arises from the derivations in [13]. We are now in a position to derive the necessary majorizers for the construction of our IRLS algorithms.

## 3 Construction of MM Algorithms

### 3.1 Derivations of Majorizers

We begin by majorizing the mentioned loss functions from the introduction. We shall denote the loss functions by $H$ (hinge loss), $LS$ (least-square), $S$ (squared-hinge), and $L$ (logistic). Note that the parameter in each of the loss functions is $w$.

We can apply Corollary 5 directly to the hinge loss $l_H(w, y) = [1 - wy]_+$ to get a majorizer at $v$:

$$M_H(w; v) = \frac{1}{4|1 - vy|} (1 - wy + |1 - vy|)^2.$$

Just as simply, we note that the least-squares loss $l_{LS}(w, y) = (1 - wy)^2$ is majorized by itself; that is

$$M_{LS}(w; v) = (1 - wy)^2,$$

for any $v$.

In order to derive a majorizer for the squared-hinge loss $l_S(w,y) = [1-wy]_+^2$, we start by writing $[u]_+^2 = (|u|/2 + u/2)^2 = u^2/2 + u|u|/2$. Noting that this simply implies that the function is identical to $u^2$ when $u \geq 0$, and 0 otherwise, we can majorize the function for any $v$ in the two separate domains of behavior via the joint function

$$M(u;v) = u^2 I(v \geq 0) + (u-v)^2 I(v < 0).$$

We obtain the desired majorizer for $l_S$ by setting $M_S(w;v) = M(1-wy; 1-vy)$, using the expression above. Here, $I(A)$ is the indicator function, which takes value 1 if proposition $A$ is true, and 0 otherwise.

Finally, to majorize the logistic loss $l_L(w,y) = \log[1 + \exp(-wy)]$, we note that the function $F(u) = \log[1 + \exp(-u)]$ has first and second derivatives $F'(u) = -\pi(u)$ and $F''(u) = \pi(u)[1 - \pi(u)]$, where $\pi(u) = \exp(-x)/[1 + \exp(-x)]$. Now note that $0 < \pi(u) < 1$ and thus $F''(u) \leq 1/4$, by an elementary quadratic maximization. Thus, we can set $\Delta = 1/4$ in Lemma 3 and majorize $F$ at $v$ by

$$M(u;v) = F(v) - \pi(v)(u-v) + \frac{(u-v)^2}{8}.$$

We obtain the desired majorizer for $l_L$ by setting $M_L(w;v) = M(1-wy; 1-vy)$, using the expression above.

We now move on to majorizing the penalty functions $P$. We shall denote the penalties $L2$ (2-norm), $L1$ (1-norm), and $E$ (elastic net). Starting with the 2-norm, as with the least-squares loss, the penalty is a majorizer of itself at every $v \in \mathbb{R}^q$. That is, $P_{L2}(\boldsymbol{\beta}) = \lambda \boldsymbol{\beta}^\top \boldsymbol{\beta}$ is majorized at any $v$ by $M_{L2}(\boldsymbol{\beta}; v) = \lambda \boldsymbol{\beta}^\top \boldsymbol{\beta} = \lambda \boldsymbol{\theta}^\top \bar{\mathbf{I}} \boldsymbol{\theta}$, where $\bar{\mathbf{I}} = \text{diag}(0, 1, ..., 1)$.

Next, the 1-norm penalty $P_{L1}(\boldsymbol{\beta}) = \mu \sum_{j=1}^q |\beta_j|$ can be majorized by appli-cation of Lemma 2 ($a = 1$) for each $j \in [q]$ ($[q] = \{1, ..., q\}$). That is, for each $j \in [q]$, we can majorize $|\beta_j|$ at $v \neq 0$ by $M_j(\beta_j; v_j) = \beta_j^2/2|v_j| + |v_j|/2$. Thus, we can write the majorizer of $P_{L1}$ as

$$M_{L1}(\boldsymbol{\beta}; \boldsymbol{v}) = \frac{\mu}{2} \sum_{j=1}^{q} \frac{\beta_j^2}{|v_j|} + \frac{\mu}{2} \sum_{j=1}^{q} |v_j|.$$

In the interest of numerical stability, we shall consider an approximation of $M_{L1}$ instead, where the denominators that are in danger of going to zero are bounded away. Let $\epsilon > 0$ be a small constant (i.e. $\epsilon = 10^{-6}$); we can approximate $M_{L1}$ by

$$M_{L1}^{\epsilon}(\boldsymbol{\beta}; \boldsymbol{v}) = \frac{\mu}{2} \sum_{j=1}^{q} \frac{\beta_j^2}{\sqrt{v_j^2 + \epsilon}} + \frac{\mu}{2} \sum_{j=1}^{q} \sqrt{v_j^2 + \epsilon}.$$

For further convenience, moving forward, we shall also write $M_{L1}^{\epsilon}$ in matrix notation with respect to the vector $\boldsymbol{\theta}$. That is, set

$$\boldsymbol{\Omega}(\boldsymbol{v}) = \mathrm{diag}\left(0, \frac{1}{\sqrt{v_j^2 + \epsilon}}, ..., \frac{1}{\sqrt{v_j^2 + \epsilon}}\right)$$

and write

$$M_{L1}^{\epsilon}(\boldsymbol{\beta}; \boldsymbol{v}) = \frac{\mu}{2} \boldsymbol{\theta}^{\top} \boldsymbol{\Omega}(\boldsymbol{v}) \boldsymbol{\theta} + \frac{\mu}{2} \sum_{j=1}^{q} \sqrt{v_j^2 + \epsilon}. \qquad (3)$$

With (3) in hand, it is now a simple process of combining the majorizers for the 1-norm and 2-norm penalties to get a majorizer for the elastic net penalty $P_E(\boldsymbol{\beta}) = \lambda \boldsymbol{\beta}^{\top} \boldsymbol{\beta} + \mu \sum_{j=1}^{q} |\beta_j|$. That is, we can majorize $P_E$ when $v_j \neq 0$ by

$$M_E(\boldsymbol{\beta}; \boldsymbol{v}) = \lambda \boldsymbol{\beta}^{\top} \boldsymbol{\beta} + \frac{\mu}{2} \sum_{j=1}^{q} \frac{\beta_j^2}{|v_j|} + \frac{\mu}{2} \sum_{j=1}^{q} |v_j|,$$

which we can approximate by

$$M_E^\epsilon(\boldsymbol{\beta}; \boldsymbol{v}) = \lambda \boldsymbol{\theta}^\top \bar{\mathbf{I}} \boldsymbol{\theta} + \frac{\mu}{2} \boldsymbol{\theta}^\top \boldsymbol{\Omega}(\boldsymbol{v}) \boldsymbol{\theta} + \frac{\mu}{2} \sum_{j=1}^q \sqrt{v_j^2 + \epsilon}.$$

## 3.2 IRLS Algorithms

### 3.2.1 Hinge Loss

We can now construct our first IRLS algorithm for fitting SVMs. We begin with the original setup of [3] (i.e. $R_n = n^{-1} l_H + P_{L2}$). Let $\mathbf{z} = \{\boldsymbol{z}_i\}_{i=1}^n$ be a fixed observation of the sample $\mathbf{Z}$. We can write the risk of the sample as

$$R_n(\alpha, \boldsymbol{\beta}; \mathbf{z}) = \frac{1}{n} \sum_{i=1}^n l_H(\alpha + \boldsymbol{\beta}^\top \boldsymbol{t}_i, y_i) + P_{L2}(\boldsymbol{\beta}).$$

We can majorize $R_n$ at some $\boldsymbol{\theta}^{(k)\top} = (\alpha^{(k)}, \boldsymbol{\beta}^{(k)\top}) \in \mathbb{R}^{q+1}$ by

$$M(\boldsymbol{\theta}; \boldsymbol{\theta}^*) = \frac{1}{n} \sum_{i=1}^n \frac{\left[1 - w(\boldsymbol{\theta}; \boldsymbol{t}_i) y_i + \gamma(\boldsymbol{\theta}^{(k)}; \boldsymbol{z}_i)\right]^2}{4\gamma(\boldsymbol{\theta}^{(k)}; \boldsymbol{z}_i)} + \lambda \boldsymbol{\theta}^\top \bar{\mathbf{I}} \boldsymbol{\theta},$$

where $w(\boldsymbol{\theta}; \boldsymbol{t}_i) = \alpha + \boldsymbol{\beta}^\top \boldsymbol{t}_i$ and $\gamma(\boldsymbol{\theta}; \boldsymbol{z}_i) = |1 - w(\boldsymbol{\theta}; \boldsymbol{t}_i) y_i|$, for each $i \in [n]$. As with the case of $M_{L1}$, there is a potential division by zero here. Thus, we shall approximate $M$ by $M^\epsilon$, where

$$M^\epsilon(\boldsymbol{\theta}; \boldsymbol{\theta}^*) = \frac{1}{n} \sum_{i=1}^n \frac{\left[1 - w(\boldsymbol{\theta}; \boldsymbol{t}_i) y_i + \gamma^\epsilon(\boldsymbol{\theta}^{(k)}; \boldsymbol{z}_i)\right]^2}{4\gamma^\epsilon(\boldsymbol{\theta}^{(k)}; \boldsymbol{z}_i)} + \lambda \boldsymbol{\theta}^\top \bar{\mathbf{I}} \boldsymbol{\theta},$$

and $\gamma^\epsilon(\boldsymbol{\theta}; \boldsymbol{z}_i) = \sqrt{(1 - w(\boldsymbol{\theta}; \boldsymbol{t}_i) y_i)^2 + \epsilon}$, for a small $\epsilon > 0$.

Following the derivation of [18], we rearrange $M^\epsilon$ and write

$$M^\epsilon\left(\boldsymbol{\theta}; \boldsymbol{\theta}^{(k)}\right) = \frac{1}{n}\left(\boldsymbol{\gamma}^{(k)} - \mathbf{Y}\boldsymbol{\theta}\right)^\top \mathbf{W}^{(k)}\left(\boldsymbol{\gamma}^{(k)} - \mathbf{Y}\boldsymbol{\theta}\right) \\ + \lambda\boldsymbol{\theta}^\top \bar{\mathbf{I}}\boldsymbol{\theta}, \qquad (4)$$

where $\mathbf{Y}^\top \in \mathbb{R}^{(q+1)\times n}$ has rows $\boldsymbol{y}_i^\top = (y_i, y_i \boldsymbol{t}_i^\top)$,

$$\boldsymbol{\gamma}^{(k)\top} = \left(\gamma^\epsilon\left(\boldsymbol{\theta}^*; \boldsymbol{z}_1\right) + 1, ..., \gamma^\epsilon\left(\boldsymbol{\theta}^{(k)}; \boldsymbol{z}_n\right) + 1\right), \text{ and}$$

$$\mathbf{W}^{(k)} = \mathrm{diag}\left(\frac{1}{4\gamma^\epsilon\left(\boldsymbol{\theta}^{(k)}; \boldsymbol{z}_1\right)}, ..., \frac{1}{4\gamma^\epsilon\left(\boldsymbol{\theta}^{(k)}; \boldsymbol{z}_n\right)}\right).$$

From (4), it is easy to see that the majorizer is in a positive-quadratic form. Thus, a global minimum can be found by solving the first-order condition (FOC) of (4) (with respect to $\boldsymbol{\theta}$) to obtain

$$\boldsymbol{\theta}^* = \left(\mathbf{Y}^\top \mathbf{W}^{(k)} \mathbf{Y} + n\lambda \bar{\mathbf{I}}\right)^{-1} \mathbf{Y}^\top \mathbf{W}^{(k)} \boldsymbol{\gamma}^{(k)}. \qquad (5)$$

We can then use (5) to construct the IRLS algorithm $\boldsymbol{\theta}^{(k+1)} = \boldsymbol{\theta}^*$ for the computation of (1) in the original setup of [3].

Starting from (4) and using (3), we can write an approximate majorizer for the risk combination of hinge loss and 1-norm penalty (i.e. $R_n = n^{-1}\sum l_H + P_{L1}$) at $\boldsymbol{\theta}^{(k)}$ as

$$M^\epsilon\left(\boldsymbol{\theta}; \boldsymbol{\theta}^{(k)}\right) = \frac{1}{n}\left(\boldsymbol{\gamma}^{(k)} - \mathbf{Y}\boldsymbol{\theta}\right)^\top \mathbf{W}^{(k)}\left(\boldsymbol{\gamma}^{(k)} - \mathbf{Y}\boldsymbol{\theta}\right) \\ + \frac{\mu}{2}\boldsymbol{\theta}^\top \boldsymbol{\Omega}^{(k)}\boldsymbol{\theta} + c, \qquad (6)$$

where $c$ is a constant that does not involve $\boldsymbol{\theta}$ and $\boldsymbol{\Omega}^{(k)} = \boldsymbol{\Omega}\left(\boldsymbol{\beta}^{(k)}\right)$. Solving the

FOC of (6) yields

$$\boldsymbol{\theta}^* = \left(\mathbf{Y}^\top \mathbf{W}^{(k)} \mathbf{Y} + n\frac{\mu}{2}\boldsymbol{\Omega}^{(k)}\right)^{-1} \mathbf{Y}^\top \mathbf{W}^{(k)} \boldsymbol{\gamma}^{(k)}. \quad (7)$$

Thus, using (7), $\boldsymbol{\theta}^{(k+1)} = \boldsymbol{\theta}^*$ defines an IRLS algorithm for computing (1) in the $R_n = n^{-1}\sum l_H + P_{L1}$ case.

Lastly, for the hinge loss variants, in the case of the risk combination of hinge loss and elastic net penalty (i.e. $R_n = n^{-1}\sum l_H + P_E$), we can approximate the majorizer by

$$\begin{aligned} M^\epsilon\left(\boldsymbol{\theta}; \boldsymbol{\theta}^{(k)}\right) &= \frac{1}{n}\left(\boldsymbol{\gamma}^{(k)} - \mathbf{Y}\boldsymbol{\theta}\right)^\top \mathbf{W}^{(k)} \left(\boldsymbol{\gamma}^{(k)} - \mathbf{Y}\boldsymbol{\theta}\right) \\ &\quad + \lambda \boldsymbol{\theta}^\top \bar{\mathbf{I}} \boldsymbol{\theta} + \frac{\mu}{2}\boldsymbol{\theta}^\top \boldsymbol{\Omega}^{(k)} \boldsymbol{\theta} + c. \end{aligned} \quad (8)$$

Again, solving the FOC yields the minimizer

$$\boldsymbol{\theta}^* = \left(\mathbf{Y}^\top \mathbf{W}^{(k)} \mathbf{Y} + n\lambda\bar{\mathbf{I}} + n\frac{\mu}{2}\boldsymbol{\Omega}^{(k)}\right)^{-1} \mathbf{Y}^\top \mathbf{W}^{(k)} \boldsymbol{\gamma}^{(k)}, \quad (9)$$

which can be used to define the IRLS algorithm $\boldsymbol{\theta}^{(k+1)} = \boldsymbol{\theta}^*$ for computing (1), in this case.

### 3.2.2 Least-Squares Loss

We consider the easier cases of combinations involving the least-squares loss $l_{LS}$. When combined with the 2-norm penalty, we can write the risk as

$$\begin{aligned} R_n(\boldsymbol{\alpha}, \boldsymbol{\beta}; \mathbf{z}) &= \frac{1}{n}\sum_{i=1}^n \left(1 - w\left(\boldsymbol{\theta}; \boldsymbol{t}_i\right) y_i\right)^2 + \lambda \boldsymbol{\theta}^\top \bar{\mathbf{I}} \boldsymbol{\theta} \\ &= \frac{1}{n}(\mathbf{1} - \mathbf{Y}\boldsymbol{\theta})^\top (\mathbf{1} - \mathbf{Y}\boldsymbol{\theta}) + \lambda \boldsymbol{\theta}^\top \bar{\mathbf{I}} \boldsymbol{\theta}, \end{aligned}$$

where **1** is a vector of ones. Since the risk is already in a quadratic form, there is no need for an IRLS algorithm, and (1) can be obtained by solving the FOC, which yields the solution

$$\hat{\boldsymbol{\theta}} = \left(\mathbf{Y}^\top \mathbf{Y} + n\lambda \bar{\mathbf{I}}\right)^{-1} \mathbf{Y}^\top \mathbf{1}.$$

In the cases where we combine $l_{LS}$ with either $P_{L1}$ or $P_E$, however, we are required to use IRLS schemes due to the forms of the penalties. For the risk combination $R_n = n^{-1} \sum l_{LS} + P_{L1}$, the approximate majorizer at $\boldsymbol{\theta}^{(k)}$ is

$$M^\epsilon\left(\boldsymbol{\theta}; \boldsymbol{\theta}^{(k)}\right) = \frac{1}{n}\left(\mathbf{1} - \mathbf{Y}\boldsymbol{\theta}\right)^\top \left(\mathbf{1} - \mathbf{Y}\boldsymbol{\theta}\right) + \frac{\mu}{2}\boldsymbol{\theta}^\top \boldsymbol{\Omega}^{(k)} \boldsymbol{\theta} + c.$$

The solution to the FOC is

$$\boldsymbol{\theta}^* = \left(\mathbf{Y}^\top \mathbf{Y} + n\frac{\mu}{2}\boldsymbol{\Omega}^{(k)}\right)^{-1} \mathbf{Y}^\top \mathbf{1}, \tag{10}$$

and thus the IRLS algorithm for computing (1) is to set $\boldsymbol{\theta}^{(k+1)} = \boldsymbol{\theta}^*$. Similarly, the IRLS algorithm for computing (1) in the $R_n = n^{-1} \sum l_{LS} + P_E$ case is to set $\boldsymbol{\theta}^{(k+1)} = \boldsymbol{\theta}^*$, where

$$\boldsymbol{\theta}^* = \left(\mathbf{Y}^\top \mathbf{Y} + n\lambda \bar{\mathbf{I}} + n\frac{\mu}{2}\boldsymbol{\Omega}^{(k)}\right)^{-1} \mathbf{Y}^\top \mathbf{1}. \tag{11}$$

### 3.2.3 Squared-Hinge Loss

Next, we consider the combinations involving the squared-hinge loss $l_S$. For the combination $R_n = n^{-1} \sum l_S + P_{L2}$, we can majorize $R_n$ at $\boldsymbol{\theta}^{(k)}$ by $M\left(\boldsymbol{\theta}; \boldsymbol{\theta}^{(k)}\right)$, which equals

$$\frac{1}{n}\sum_{i=1}^{n}\left(1 - w\left(\boldsymbol{\theta}; \boldsymbol{t}_i\right) y_i\right)^2 \left[1 - v\left(\boldsymbol{\theta}^{(k)}; \boldsymbol{z}_i\right)\right]$$

$$+\frac{1}{n}\sum_{i=1}^{n}\left(1-w\left(\boldsymbol{\theta};\boldsymbol{t}_i\right)y_i-\delta\left(\boldsymbol{\theta}^{(k)};\boldsymbol{z}_i\right)\right)^2 \upsilon\left(\boldsymbol{\theta}^{(k)};\boldsymbol{z}_i\right)$$
$$+\lambda\boldsymbol{\theta}^\top\bar{\mathbf{I}}\boldsymbol{\theta},$$

where $\delta\left(\boldsymbol{\theta}^{(k)};\boldsymbol{z}_i\right) = 1 - w\left(\boldsymbol{\theta}^{(k)};\boldsymbol{t}_i\right)y_i$, and $\upsilon\left(\boldsymbol{\theta}^{(k)};\boldsymbol{z}_i\right) = I\left(\delta\left(\boldsymbol{\theta}^{(k)};\boldsymbol{z}_i\right) < 0\right)$. We can rewrite $M$ as

$$\begin{aligned}
M\left(\boldsymbol{\theta};\boldsymbol{\theta}^{(k)}\right) &= \frac{1}{n}(\mathbf{1}-\mathbf{Y}\boldsymbol{\theta})^\top\left[\mathbf{I}-\boldsymbol{\Upsilon}^{(k)}\right](\mathbf{1}-\mathbf{Y}\boldsymbol{\theta})\\
&+\frac{1}{n}\left(\boldsymbol{\delta}^{(k)}-\mathbf{Y}\boldsymbol{\theta}\right)^\top\boldsymbol{\Upsilon}^{(k)}\left(\boldsymbol{\delta}^{(k)}-\mathbf{Y}\boldsymbol{\theta}\right)\\
&+\lambda\boldsymbol{\theta}^\top\bar{\mathbf{I}}\boldsymbol{\theta},
\end{aligned} \quad (12)$$

where $\mathbf{I}$ is the identity matrix,

$$\boldsymbol{\delta}^{(k)\top} = \left(1-\delta\left(\boldsymbol{\theta}^{(k)};\boldsymbol{z}_1\right),...,1-\delta\left(\boldsymbol{\theta}^{(k)};\boldsymbol{z}_n\right)\right), \text{ and}$$

$$\boldsymbol{\Upsilon}^{(k)} = \text{diag}\left(\upsilon\left(\boldsymbol{\theta}^{(k)};\boldsymbol{z}_1\right),...,\upsilon\left(\boldsymbol{\theta}^{(k)};\boldsymbol{z}_n\right)\right).$$

Solving the FOC for (12) yields the solution

$$\begin{aligned}
\boldsymbol{\theta}^* &= \left(\mathbf{Y}^\top\mathbf{Y}+n\lambda\bar{\mathbf{I}}\right)^{-1}\mathbf{Y}^\top\left(\mathbf{I}-\boldsymbol{\Upsilon}^{(k)}\right)\mathbf{1}\\
&+\left(\mathbf{Y}^\top\mathbf{Y}+n\lambda\bar{\mathbf{I}}\right)^{-1}\mathbf{Y}^\top\boldsymbol{\Upsilon}^{(k)}\boldsymbol{\delta}^{(k)},
\end{aligned} \quad (13)$$

which we can use to define the $(k+1)$th iteration of the IRLS algorithm $\boldsymbol{\theta}^{(k+1)} = \boldsymbol{\theta}^*$ for computing (1) in this case.

From (12), it is not difficult to deduce that an approximate majorizer for the $R_n = n^{-1}\sum l_S + P_{L1}$ will take the form

$$\begin{aligned}
M^\epsilon\left(\boldsymbol{\theta};\boldsymbol{\theta}^{(k)}\right) &= \frac{1}{n}(\mathbf{1}-\mathbf{Y}\boldsymbol{\theta})^\top\left[\mathbf{I}-\boldsymbol{\Upsilon}^{(k)}\right](\mathbf{1}-\mathbf{Y}\boldsymbol{\theta})\\
&+\frac{1}{n}\left(\boldsymbol{\delta}^{(k)}-\mathbf{Y}\boldsymbol{\theta}\right)^\top\boldsymbol{\Upsilon}^{(k)}\left(\boldsymbol{\delta}^{(k)}-\mathbf{Y}\boldsymbol{\theta}\right)
\end{aligned}$$

$$+\frac{\mu}{2}\boldsymbol{\theta}^\top\boldsymbol{\Omega}^{(k)}\boldsymbol{\theta},$$

for which we can solve the FOC to obtain the solution

$$\begin{aligned}\boldsymbol{\theta}^* &= \left(\mathbf{Y}^\top\mathbf{Y}+n\frac{\mu}{2}\boldsymbol{\Omega}^{(k)}\right)^{-1}\mathbf{Y}^\top\left(\mathbf{I}-\boldsymbol{\Upsilon}^{(k)}\right)\mathbf{1} \\ &+ \left(\mathbf{Y}^\top\mathbf{Y}+n\frac{\mu}{2}\boldsymbol{\Omega}^{(k)}\right)^{-1}\mathbf{Y}^\top\boldsymbol{\Upsilon}^{(k)}\boldsymbol{\delta}^{(k)}.\end{aligned} \quad (14)$$

We can then define the IRLS for this case as $\boldsymbol{\theta}^{(k+1)} = \boldsymbol{\theta}^*$. From the previous results, we can quickly deduce that the IRLS for computing (1) in the $R_n = n^{-1}\sum l_S + P_E$ case is to take $\boldsymbol{\theta}^{(k+1)} = \boldsymbol{\theta}^*$, where

$$\begin{aligned}\boldsymbol{\theta}^* &= \left(\mathbf{Y}^\top\mathbf{Y}+n\lambda\bar{\mathbf{I}}+n\frac{\mu}{2}\boldsymbol{\Omega}^{(k)}\right)^{-1}\mathbf{Y}^\top\left(\mathbf{I}-\boldsymbol{\Upsilon}^{(k)}\right)\mathbf{1} \\ &+ \left(\mathbf{Y}^\top\mathbf{Y}+n\lambda\bar{\mathbf{I}}+n\frac{\mu}{2}\boldsymbol{\Omega}^{(k)}\right)^{-1}\mathbf{Y}^\top\boldsymbol{\Upsilon}^{(k)}\boldsymbol{\delta}^{(k)}.\end{aligned} \quad (15)$$

### 3.2.4 Logistic Loss

Finally, we consider the combinations involving the logistic loss $l_L$. The majorizer $M\left(\boldsymbol{\theta};\boldsymbol{\theta}^{(k)}\right)$ for the risk for the combination with the 2-norm penalty (i.e. $R_n = n^{-1}\sum l_L + P_{L2}$) can be written as

$$\begin{aligned}&\frac{1}{n}\sum_{i=1}^n \log\left(1+\exp\left[-w\left(\boldsymbol{\theta};\boldsymbol{t}_i\right)y_i\right]\right) \\ &-\frac{1}{n}\sum_{i=1}^n \pi_i^{(k)}\left[w\left(\boldsymbol{\theta};\boldsymbol{t}_i\right)y_i - w\left(\boldsymbol{\theta}^{(r)};\boldsymbol{t}_i\right)y_i\right] \\ &+\frac{1}{n}\sum_{i=1}^n \frac{\left[w\left(\boldsymbol{\theta};\boldsymbol{t}_i\right)y_i - w\left(\boldsymbol{\theta}^{(r)};\boldsymbol{t}_i\right)y_i\right]^2}{8} + \lambda\boldsymbol{\theta}^\top\bar{\mathbf{I}}\boldsymbol{\theta},\end{aligned}$$

where $\pi_i^{(k)} = \pi\left(w\left(\boldsymbol{\theta}^{(k)}; \boldsymbol{t}_i\right) y_i\right)$. We can rewrite the majorizer as

$$
\begin{aligned}
M\left(\boldsymbol{\theta}; \boldsymbol{\theta}^{(k)}\right) &= \frac{1}{8n}\left(\boldsymbol{\delta}^{(k)} - \mathbf{Y}\boldsymbol{\theta}\right)^\top \left(\boldsymbol{\delta}^{(k)} - \mathbf{Y}\boldsymbol{\theta}\right) \\
&\quad - \frac{1}{n}\boldsymbol{\pi}^{(k)\top}\mathbf{Y}\boldsymbol{\theta} + \lambda \boldsymbol{\theta}^\top \bar{\mathbf{I}}\boldsymbol{\theta} + c.
\end{aligned} \qquad (16)
$$

The FOC solution for (16) iS

$$
\begin{aligned}
\boldsymbol{\theta}^* &= \left(\mathbf{Y}^\top \mathbf{Y} + 8n\lambda \bar{\mathbf{I}}\right)^{-1} \mathbf{Y}^\top \boldsymbol{\delta}^{(k)} \\
&\quad + 4\left(\mathbf{Y}^\top \mathbf{Y} + 8n\lambda \bar{\mathbf{I}}\right)^{-1} \mathbf{Y}^\top \boldsymbol{\pi}^{(k)},
\end{aligned} \qquad (17)
$$

which yields the IRLS $\boldsymbol{\theta}^{(k+1)} = \boldsymbol{\theta}^*$ for the computation of (1).

As with the previous loss functions, we must approximate the majorizer for the case of the 1-norm penalty. Thus, an approximate majorizer for the case $R_n = n^{-1}\sum l_L + P_{L1}$ can be written as

$$
\begin{aligned}
M^\epsilon\left(\boldsymbol{\theta}; \boldsymbol{\theta}^{(k)}\right) &= \frac{1}{8n}\left(\boldsymbol{\delta}^{(k)} - \mathbf{Y}\boldsymbol{\theta}\right)^\top \left(\boldsymbol{\delta}^{(k)} - \mathbf{Y}\boldsymbol{\theta}\right) \\
&\quad - \frac{1}{n}\boldsymbol{\pi}^{(k)\top}\mathbf{Y}\boldsymbol{\theta} + \frac{\mu}{2}\boldsymbol{\theta}^\top \boldsymbol{\Omega}^{(k)}\boldsymbol{\theta} + c,
\end{aligned}
$$

with the corresponding FOC solution

$$
\begin{aligned}
\boldsymbol{\theta}^* &= \left(\mathbf{Y}^\top \mathbf{Y} + 4n\mu \boldsymbol{\Omega}^{(k)}\right)^{-1} \mathbf{Y}_n^\top \boldsymbol{\delta}^{(k)} \\
&\quad + 4\left(\mathbf{Y}^\top \mathbf{Y} + 4n\mu \boldsymbol{\Omega}^{(k)}\right)^{-1} \mathbf{Y}^\top \boldsymbol{\pi}^{(k)},
\end{aligned} \qquad (18)
$$

which leads to the IRLS algorithm $\boldsymbol{\theta}^{(k+1)} = \boldsymbol{\theta}^*$.

Lastly, from the solution (18), it is not difficult to deduce that the IRLS for computing (1) in the $R_n = n^{-1}\sum l_L + P_E$ case is to take $\boldsymbol{\theta}^{(k+1)} = \boldsymbol{\theta}^*$, where

$$
\boldsymbol{\theta}^* = \left(\mathbf{Y}^\top \mathbf{Y} + 8n\lambda \bar{\mathbf{I}} + 4n\mu \boldsymbol{\Omega}^{(k)}\right)^{-1} \mathbf{Y}^\top \boldsymbol{\delta}^{(k)} \qquad (19)
$$

$$+4\left(\mathbf{Y}^\top\mathbf{Y} + 8n\lambda\bar{\mathbf{I}} + 4n\mu\mathbf{\Omega}^{(k)}\right)^{-1}\mathbf{Y}^\top\boldsymbol{\pi}^{(k)}.$$

## 3.3 Some Theoretical Results

In every combination except for the least-squares loss with 2-norm penalty, the MM algorithm that is derived is an IRLS algorithm. In the cases of the squared-hinge and logistic losses, in combination with the 2-norm penalty, the derived IRLS algorithm minimizes exact majorizers of the risk functions corresponding to the respective combinations. As such, Propositions 1 and 2 along with the quadratic forms of each of the risks yield the following result.

**Theorem 6.** *If $\{\boldsymbol{\theta}^{(k)}\}$ is a sequence that is obtained via the IRLS algorithm defined by the iterations $\boldsymbol{\theta}^{(k+1)} = \boldsymbol{\theta}^*$, where $\boldsymbol{\theta}^*$ takes the form (13) or (17), then the sequence of risk values $\{R_n(\alpha^{(k)}, \boldsymbol{\beta}^{(k)}; \mathbf{z})\}$ monotonically decreases, where $R_n$ takes the form $n^{-1}\sum l_S + P_{L2}$ or $n^{-1}\sum l_L + P_{L2}$, respectively. Furthermore, if $\boldsymbol{\theta}^{(\infty)}$ is the limit point of either IRLS algorithms, then $\boldsymbol{\theta}^{(\infty)}$ is a global minimizer of the respective risk function.*

It may serve as a minor source of dissatisfaction that in all other combinations (i.e. combinations involving either the hinge loss or the 1-norm and elastic net penalties) approximations to majorizers of the respective risk functions must be made. As such, although practically rare, the sequences of IRLS algorithm iterates for each of the aforementioned cases are not guaranteed to monotonically decrease the respective risk functions that they were derived from. However, this is not to say that the approximations are arbitrary as we shall see from the following basic calculus fact (cf. [12, Eqn. 4.7]).

**Lemma 7.** *If $F(u) = f(u)$ is a concave and differentiable function for some $u \in \mathbb{U}$, then $F$ is majorized at any $v \in \mathbb{U}$ by*

$$M(u; v) = f(v) + f'(v)(u - v).$$

**Corollary 8.** *The function* $F(u) = \sqrt{u}$ *is majorized at any* $v \geq 0$ *by*

$$M(u;v) = \sqrt{v} + (u-v)/(2\sqrt{y}).$$

Applying Corollary 8 with $F(u^2 + \epsilon)$ to majorize at some $v^2 + \epsilon$, where $\epsilon > 0$, yields the majorizer

$$M(u^2 + \epsilon; v^2 + \epsilon) = \sqrt{v^2 + \epsilon} + \frac{(u^2 - v^2)}{2\sqrt{v^2 + \epsilon}}. \tag{20}$$

We note that as $\epsilon$ approaches 0, $\sqrt{u^2 + \epsilon} \to |u|$ uniformly in $u$. Furthermore, (20) is precisely the form of the approximations that are used when the risk function involves either the hinge loss or the 1-norm and elastic net penalties. Thus, we can approximate any occurrence of an absolute value function by $\sqrt{u^2 + \epsilon}$. As such, the following result is available.

**Theorem 9.** *If* $\{\boldsymbol{\theta}^{(k)}\}$ *is a sequence that is obtained via the IRLS algorithm defined by the iterations* $\boldsymbol{\theta}^{(k+1)} = \boldsymbol{\theta}^*$, *where* $\boldsymbol{\theta}^*$ *takes the forms (7), (9), (10), (15), (14), (15), (18), or (19), then there exists an approximate risk function* $R_n^\epsilon(\boldsymbol{\alpha}, \boldsymbol{\beta}; \mathbf{z})$, *for which the sequence* $\{R_n^\epsilon(\boldsymbol{\alpha}^{(k)}, \boldsymbol{\beta}^{(k)}; \mathbf{z})\}$ *is monotonically decreasing, where* $R_n^\epsilon(\boldsymbol{\alpha}, \boldsymbol{\beta}; \mathbf{z})$ *uniformly converges to* $R_n(\boldsymbol{\alpha}, \boldsymbol{\beta}; \mathbf{z})$ *(as* $\epsilon$ *approaches 0) and* $R_n$ *is the risk function corresponding to the combination of loss and penalty of each algorithm. Furthermore, if* $\boldsymbol{\theta}^{(\infty)}$ *is a limit point of the respective algorithm, then* $\boldsymbol{\theta}^{(\infty)}$ *is a global minimizer of the respective approximate risk* $R_n^\epsilon$.

*Remark* 10. In both the cases where $R_n$ is minorized by $M_n$ or $R_n^\epsilon$ by $M_n^\epsilon$, the pairs of risk and majorizer functions are continuous and differentiable, respectively (for fixed $\epsilon$ in the approximate cases). Thus, in both cases, we can apply [21, Prop. 1], which establishes the satisfaction of the Proposition 2 hypotheses for differentiable functions. The results of Theorems 6 and 9 then follow from the MM construction of the algorithms.

*Remark* 11. We note that approximations of objectives or majorizers are inevitable in the construction of MM algorithms for non-differentiable optimization problem. See, for example, the similar approaches that are taken in [8] and [10].

## 4 Numerical Examples

To demonstrate the properties of the IRLS algorithms, we now conduct a numerical study involving a small simulation. Let $\mathbf{z}_n$ be a realization of a sample from the following process. Fix $n = 10000$, for $i \in [5000]$, we set $Y_i = -1$ and simulate $\boldsymbol{X}_i \in \mathbb{R}^2$ from a spherical normal distribution with mean vector $(-1, -1)$. Similarly, for $i \in [n] \setminus [5000]$, we set $Y_i = 1$ and simulate $\boldsymbol{X}_i$ from a spherical normal distribution with mean vector $(1, 1)$. No transformation is made and so we set $\boldsymbol{T}_i = \boldsymbol{X}_i$, for each $i \in [n]$.

Using the simulated realization, we ran algorithms (13) and (17) for 50 iterations each, to fit SVMs with risk functions $R_n = n^{-1} \sum l_S + P_{L2}$ and $R_n = n^{-1} \sum l_L + P_{L2}$. The risk trajectories of the two algorithms for various values of $\lambda$ are visualized in Figures 1 and 2, respectively. We also ran algorithms (6) and (10) for 50 iterations each, to fit SVMs with risk functions $R_n = n^{-1} \sum l_H + P_{L1}$ and $R_n = n^{-1} \sum l_{LS} + P_{L1}$. The risk trajectories of the two algorithms for various values of $\mu$ are visualized in Figures 3 and 4, respectively.

In order to verify that all of the algorithms are working as intended, we also provide a visualization of some separating hyperplanes. Figure 5 displays the separating hyperplanes for the four algorithms that are presented above with penalization constants set at $\lambda = 0.4$ or $\mu = 0.4$.

We observe that that the monotonicity guarantees of Theorem 6 are realized in Figures 1 and 2. Furthermore, in Figures 3 and 4, we observe that even for algorithms (6) and (10), where there are no monotonicity guarantees with respect to the actual risk (and not the approximate risk of Theorem 9), the risk trajectories are still monotonically decreasing. Thus, we can conclude with some confidence that the approximations to the majorizers that are made in order to construct the IRLS algorithms, where necessary, do not appear to affect the performances of the IRLS algorithms with respect to the minimization of the respective risk functions. Upon inspection of Figure 5, we see that all four tested algorithms appear to generate hyperplanes that separate the labelled samples well.

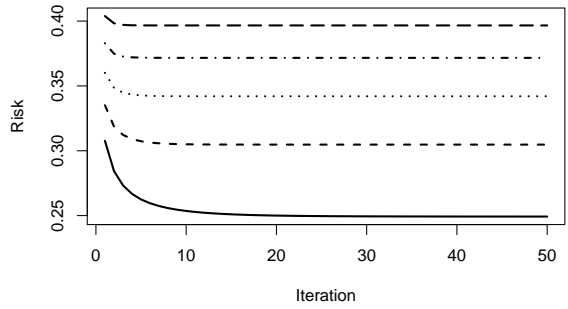

Figure 1: Risk function sequences corresponding to algorithm (13). Solid, dashed, dotted, dashed-dotted, and long dashed curves represent values $\lambda = 0, 0.1, 0.2, 0.3, 0.4$, respectively.

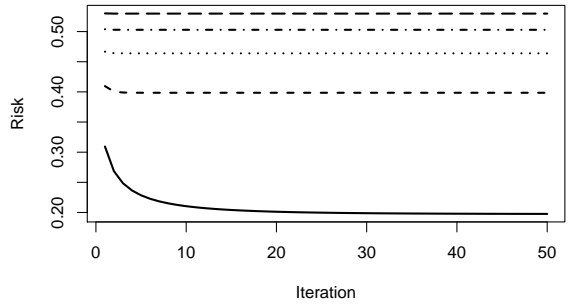

Figure 2: Risk function sequences corresponding to algorithm (17). Solid, dashed, dotted, dashed-dotted, and long dashed curves represent values $\lambda = 0, 0.1, 0.2, 0.3, 0.4$, respectively.

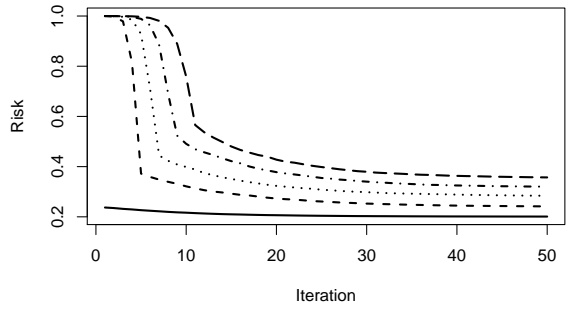

Figure 3: Risk function sequences corresponding to algorithm (6). Solid, dashed, dotted, dashed-dotted, and long dashed curves represent values $\mu = 0, 0.1, 0.2, 0.3, 0.4$, respectively.

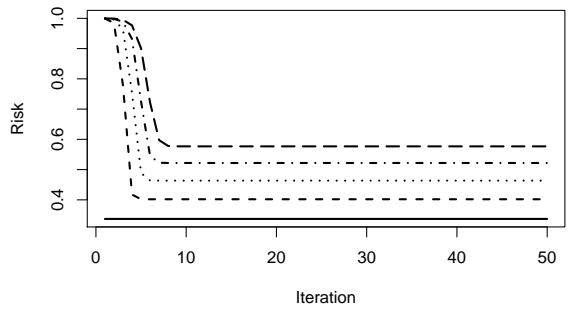

Figure 4: Risk function sequences corresponding to algorithm (10). Solid, dashed, dotted, dashed-dotted, and long dashed curves represent values $\mu = 0, 0.1, 0.2, 0.3, 0.4$, respectively.

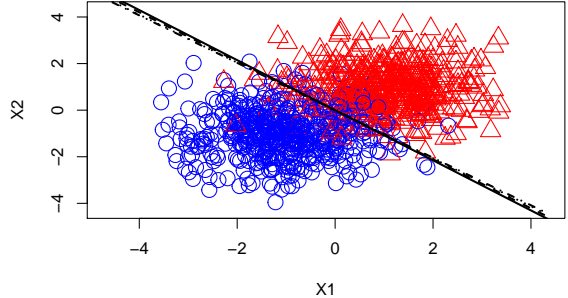

Figure 5: Separating hyperplanes obtained via four different SVM risk combinations. Solid, dashed, dotted, and dashed-dotted lines represent the hinge, least-square, squared-hinge, and logistic loss hyperplanes, respectively. Circles and triangles represent observations with label $Y_i = -1$ and $Y_i = 1$, respectively.

## 5 Conclusions

We have demonstrated that numerous SVM risk combinations, involving different losses and penalties, can be minimized using IRLS algorithms that are constructed via the MM paradigm. Although we have only assessed four loss functions in this article, there are numerous other currently used losses that can be minimized via MM-derived IRLS algorithms. For example, we have identified the quadratically-smoothed loss, Huber loss, and modified Huber loss of [30] as well as the Huberized-hinge loss of [29] as potential candidates for IRLS implementations.

If one wishes to undertake distributed or parallel estimation of SVMs, then the IRLS algorithms that we have presented fit easily with the distributed framework that is developed for the [3] specification by [17]. Furthermore, under appropriate probabilistic assumptions on the data generating process of $\mathbf{Z}$, the software alchemy approach of [15] for embarrassingly parallel problems can be applied in order to combine locally-computed SVM fits to yield a combined fit with the same statistical property as a fit on all of the locally-distributed data, simultaneously; see also [14].

As we mentioned in the introduction, the presented methodology is currently experimental and illustrative of what is possible when one seeks to estimate SVM classifiers using MM algorithm techniques for optimization. The presented implementations are not, in their current forms, replacements for the respective state-of-the-art fitting techniques, such as those implemented in the packages referenced in [23]. However, we believe that with refinement and hybridization with current best-practice approaches, there may be avenues for potential computational gains on the current state-of-the-art via MM algorithm techniques. Furthermore, MM algorithms may also be useful for estimation of SVMs with loss and penalty combinations for which best-practice currently do not exist.

We note that a minor caveat to our approach is the lack of ability to utilize the kernel trick for implicit feature space mapping. However, we believe that with a well-chosen transformation space $\mathbb{T}$, this should be easily resolved. Overall, we believe that the MM paradigm presents a novel perspective to algorithm construction for SVMs and has the potential to be used in interesting and productive ways.

## Acknowledgements

HDN is funded by Australian Research Council grant DE170101134. GJM is funded by a Discovery Projects grant from the Australian Research Council.